\documentclass[a4paper,12pt]{article}
\begin{document}
\begin{center}
\begin{Large}
\bf{Optical quantum computer based on RDS crystal}\\
\end{Large}
\bigskip
Z. Sazonova\\
Physics Department, Automobile $\&$ Road-Construction Institute (Technical University)
64, Leningradskii prospect, Moscow 125829, Russia\\
\bigskip

R. Singh\\
Wave Research Center at General Physics Institute of Russian Academy of Sciences,
38, Vavilov street, Moscow 117942, Russia Tel./Fax: (+7 095) 135-8234 email: ranjit@dataforce.net \\

\begin{abstract}
We have proposed the construction of optical quantum computer (OQC) on regular 
domain structure (RDS) crystal. By using RDS crystal, we can perform all the
logical operations on one RDS crystal. Moreover, RDS crystals are parctically
independent to the heating effects [4] i.e., can perform logic operations constantly 
without cooling the RDS crystal [4]. Also, we have proposed the quantum parallelsim 
i.e., parallel coherent laser beams are injected at the input of the RDS 
crystals. By using the RDS crystal we can perform the reduce the requirements
of the linear and nonlinear optical components.
\end{abstract}
\end{center}

\section{Introduction}
The physics of quantum computation and information [2] is one of the branch of 
quantum theory, which develops day by day very fast. A lot of proposals [2] have 
been proposed for the construction of quantum computer. Most reliable, some of 
them [2] are optical quantum computer (OQC), ion-trap, and solid-state quantum computer 
(SSQC). The most prospective technique to construct quantum computer among
them is SSQC. So, througuhout this paper, we will discuss the OQC and SSQC. 
Our dream to see the power, possibilities and 
applications of quantum computer is still under process. The reason for 
comparing the OQC with SSQC is that the coherent optical computers [1,6,7] 
which come around 1960 were studied very extensively in USA for their 
millitary purposes. The coherent optical computers were constructed with 
the huge amount of optical instruments and were useful for very limited 
number of problems. After the invention of the MASER, the requirement [1]
for coherent optical computer starts increasing very fast with more requirements of 
linear and nonlinear optical instruments. Moreover, the applications, 
possibilities of coherent optical computer were very limited as compare 
to the digital computer based on Silicon. So, the coherent optical computers 
lost their interest [6-8] due to the huge amount of requirements and limited 
number of applications. In the begining of 1990, OQC [2,3,5,9] were proposed. 
In this article, we are studying the main technical requirements to construct 
OQC and comparing with SSQC. Moreover, firstly, we are proposing to construct 
OQC based on RDS crystals [4].
 
\section{Quantum computer}
\subsection{Soild-state quantum computer (or SSQC)}
The SSQC consists [2] of $n$ number of atoms in one 
molecule. Why we are taking one molecule and why not two or three? The reason
for taking one molecule is that the interaction within the molecules are 
negligible as compare to interaction of atoms in each molecule.
Some of the atoms in each molecule have nuclear spins $1/2$ with 
slightly different resonance frequencies. To have an idea of SSQC, 
we will demonstrate the graphical representation of
SSQC and how to perform logical operations on it. As an example we are 
taking one molecule having two atoms and each atom having nuclear spin $1/2$ 
and resonance frequencies $\omega_{1}$ and $\omega_{2}$ (see fig.1). 

\subsubsection{Hamiltonian of the molecule having two nuclear spins}
The two spins are interacting with each other with hyperfine interaction $j_{12}$. 
To start, the implementation of some quantum algorithm, firstly, we align 
the two nuclear spins of each molecule along the direction of constant magnetic field $B_{0}$, 
which is applied along the $z$-axis. The Hamiltonian of the system of two 
nuclear spins $1/2$ with hyperfine interaction $j_{12}$ [2]
\begin{eqnarray}
\hat{H}=B_{0}(\hat{\sigma}_{z1}+\hat{\sigma}_{z2})+j_{12}\hat{\sigma}_{z1}\hat{\sigma}_{z2}
\end{eqnarray}
Where $\hat{\sigma}_{z1}$ and $\hat{\sigma}_{z2}$ are the Pauli matrices of 
nuclear spins $\hat{s}_{1}=\frac{1}{2}\hat{\sigma}_{1}$ and 
$\hat{s}_{2}=\frac{1}{2}\hat{\sigma}_{2}$.
The logical operations are performed by the time dependent unitary operator
\begin{eqnarray}
\hat{U}(t)=e^{-i\hat{H}t/\hbar}
\end{eqnarray}
We can change the dyanamics and position of spins $\hat{s}_{1}$ and $\hat{s}_{2}$
by selecting different resonance frequency impulses [2].
\subsubsection{Quantum logic gates}
Recently, it is shown that for the implementation of quantum algorithms on
quantum computer, only two quantum logical gates (NOT and CNOT) [2] are needed. 
It means that the complex quantum algorithms can be executed on quantum 
computer with the combination of quantum logical gates NOT and CNOT 
(control-not). The true tables for quantum logical gates NOT and CNOT 
are as follows
\begin{center}
Truth tables for logic gates NOT and CNOT:
\begin{tabular}{|c|c|c|c|c|c|c|} \cline{1-2} \cline{4-7}
\multicolumn{2}{|c|}{\slshape NOT gate} & \hspace{7mm} &
\multicolumn{4}{|c|}{\slshape CNOT gate} \\
\cline{1-2} \cline{4-7}
Input & Output & & Input & Input & Output & Output \\
$x$ & $y=\bar{x}$ & & $x_{1}$ & $x_{2}$ & $y_{1}=x_{1}$ & $y_{2}=x_{1}\oplus{x_{2}}$ \\
\cline{1-2} \cline{4-7}
0 & 0 & & 0 & 0 & 0 & 0 \\
0 & 1 & & 0 & 1 & 0 & 1 \\
  &   & & 1 & 0 & 1 & 1 \\
  &   & & 1 & 1 & 1 & 0 \\
\cline{1-2} \cline{4-7}
\end{tabular} 
\end{center}
Where bar over $\bar{x}$ and sign $\oplus$ denote the NOT operation and summation modulo $2$.
\subsubsection{Minimum requirements to construct SSQC}
\begin{enumerate}
\item $2^{n}$-number of qubits (quantum bits) are required to operate and store data.
\item To prepare the system (qubits) to its initial state i.e., constant magnetic field is required to align the initial spin states along the magnetic field direction.
\item To isolate the quantum computer (qubits) from the environment (decoherence effect) [10], which is practically uncntrollable part of the quantum computer.
\item Unitary transformations to implement quantum algorithms i.e., for the implementations of quantum algorithms.
\item Quantum measurements [10], which are needed for measuring the final state of the implemented quantum algorithms.
\end{enumerate}
\subsection{Optical quantum computer (or OQC)}
There are two proposals [2,3,5,9] to construct OQC on the basis of linear optics and 
nonlinear optics. The linear optics means that we need coherent laser beams, 
polarization beam splitters (PBS), polarization rotating plates, optical 
mirrors, photo-detectors etc. For the case of nonlinear optics we need 
nonlinear crystals and all the components of the linear optics. 
\subsubsection{Case no.1}
\subsubsection{OQC based on linear optics}
The proposal to construct OQC based on linear optics was given by Spreeuw 
[9]. To perform quantum computation on OQC for the general case, we need 
exponentially amount of linear optical components. For example to have 
$2^n$ qubits as in the case of SSQC, we need at least $2^n$ number of optical 
components to have properties like SSQC at least. The exponentially number of 
components can be reduced by using polarized states (Jones matrices) of light.
\subsubsection{OQC qubits}
The OQC $3$ qubits can be represented by
\begin{eqnarray}
|\phi>=a|0,0,0>+b|1,1,1>
\end{eqnarray}    
To have (represent) three OQC qubits (3), we need $4$ polarized light 
modes i.e., $4$ Jones matrices [9]. But in SSQC we need $n=3$ nuclear spins
to have $2^3=8$ quantum states. PBS are required for separating horizontal
and vertical polarization components of the laser mode.
\subsubsection{Why PBS?}
For performing the logical operations. For example, the logical operations 
NOT and CNOT can be implemented by using the PBS.
\subsubsection{Linear optics logical gates}
Mathematically, linear optics logical gates can be constructed by using
$SU(2)$ and $SU(1,1)$ algebra [3]. Experimentally, it can be constructed by
rotating the PBS (see fig.2).  
\subsubsection{Case no.2}
\subsubsection{OQC based on nonlinear optics}
The proposal to construct logical gates based on nonlinear optics is given by 
Milburn [6,2]. This OQC uses nonlinear crystal to get the squeezed
states of light (SSL). As we know the SSL has antibunching property i.e.,
the photons are uniformally distrubuted as compared to the coherent state
of light. By using the antibunching property of SSL, this computer can  
\subsubsection{OQC qubits}
In the case of SSL, the role of qubits perform the photons of SSL i.e., each
photon of SSL represents, one optical qubit based on nonlinear optics. 
\subsubsection{Nonlinear optics logical gates [3,5]}
To operate on each photon qubit, we need photo-detectors with high 
quantum efficieny, which are our dream of future. Moreover, in quantum 
mechanics, we detect the average number of physical quantities. It means, 
we do not have $100\%$ gurantee that we have detected the concrete photon, 
which we planned. Again, we need the beam splitters to perform the
logical operations, which are highly noisy.
\subsubsection{Case no.3}
\subsubsection{OQC based on RDS (regular domain structure) crystal}
OQC can be constructed at least on one nonlinear crystal. It means that
$n$ number of laser beams are propagating through one nonlinear crytal (see
fig. 2). Why we are using nonlinear crystal? There are some advantages to use
nonlinear crytals:
\begin{enumerate}
\item{logical gates (NOT and CNOT) can be performed by using the RDS (regular domain 
structure) crystals.}
\item{We can use the property of squeezed states of light, i.e., the quantum
noise is less as compare to the coherent state of light.}
\item{RDS crystals are independent on heating [4] of the crystals. We
can use this crystal without any cooling process, which will not delay our
computation process.}
\item{\ldots.}
\end{enumerate}
\subsubsection{Logical gates based on RDS crystal}
There are two ways to perform the logical operations in RDS crystals. 
\begin{enumerate}
\item{By using the properties of bistable switches i.e., the logical switch 
starts working when the intensity of the coherent laser beam (fundamental mode) 
propagating through the RDS crystal starts producing the predefined intensity 
of harmonic generation for some concrete logical operation.}
\item{Since, the RDS crystals can produce at a time multiple number of harmonic
modes. We can predefine each harmonic for logic operations. For example, we
can predefind second harmonic generation for NOT and third harmonic generation
for CNOT logical operations.}
\end{enumerate}
\section{Conclusion}
If we will compare the properties of SSQC and OQC, we can conclude that the 
future of OQC is practically absent. Due to the requirement of huge number 
of optical components (linear and nonlinear), heating of the linear and 
nonlinear crystals and quantum noise is greater than the coherent state of light.
The proposal which we have offered i.e., the construction of OQC on RDS crytals
require less number of components as compare to linear and nonlinear (ordinary)
crystals. Moreover, we can construct the logical gates on one RDS crytals with
lesser quantum noise, because we are performing all the logic operations in it 
as compare to the linear and nonlinear crystals [2,3,5,9]. So, the OQC based on 
RDS is more realistic as compare to other proposals to construct OQC.

Fig.1\\
Fig.2a\\
Fig.2b\\
Fig.3

\end{document}